\documentclass[twocolumn,tighten]{aastex631}
\usepackage{CJKutf8}
\usepackage{amsmath}
\usepackage{enumitem}
\usepackage{booktabs,pbox}
\usepackage{bm}

\graphicspath{{./figures/}{../figures/}}

\DeclareRobustCommand{\ion}[2]{%
\relax\ifmmode
\ifx\testbx\f@series
{\mathbf{#1\,\mathsc{#2}}}\else
{\mathrm{#1\,\mathsc{#2}}}\fi
\else\textup{#1\,{\mdseries\textsc{#2}}}%
\fi}

\def\hal [H$\alpha$]
\def\HI {\ion{H}{i}} 
\newcommand{\HII}{\ion{H}{ii}} 
\newcommand{\SII}{[\ion{S}{ii}]} 
\newcommand{\SIII}{[\ion{S}{iii}]} 

\def\2MASS {\it 2MASS}

\newcommand{\jwst}{{JWST}}
\newcommand{\hst}{{HST}}
\def\m{$\mu$m}

\shorttitle{Clusters \& PAHs}
\shortauthors{Dale/PHANGS}

\begin{document}


\title{PHANGS-JWST First Results: The Influence of Stellar Clusters on PAHs in Nearby Galaxies}



\newcommand{\UWyoming}{\affiliation{Department of Physics and Astronomy, University of Wyoming, Laramie, WY 82071, USA}}
\newcommand{\STScI}{\affiliation{Space Telescope Science Institute, 3700 San Martin Drive, Baltimore, MD 21218, USA}}
\newcommand{\UAntof}{\affiliation{Centro de Astronomía (CITEVA), Universidad de Antofagasta, Avenida Angamos 601, Antofagasta, Chile}}
\newcommand{\NOIRLab}{\affiliation{Gemini Observatory/NSF NOIRLab, 950 N. Cherry Avenue, Tucson, AZ 85719, USA}}
\newcommand{\UToledo}{\affiliation{Ritter Astrophysical Research Center, University of Toledo, Toledo, OH 43606, USA}}
\newcommand{\JHU}{\affiliation{Department of Physics and Astronomy, The Johns Hopkins University, Baltimore, MD 21218 USA}}
\newcommand{\Caltech}{\affiliation{TAPIR, California Institute of Technology, Pasadena, CA 91125 USA}}
\newcommand{\OSU}{\affiliation{Department of Astronomy, The Ohio State University, 140 West 18th Ave., Columbus, OH 43210, USA}}
\newcommand{\MPIA}{\affiliation{Max Planck Institut f\"ur Astronomie, K\"onigstuhl 17, 69117 Heidelberg, Germany}}
\newcommand{\UAlberta}{\affiliation{Department of Physics, University of Alberta, Edmonton, AB T6G 2E1, Canada}}
\newcommand{\UCSD}{\affiliation{Center for Astrophysics \& Space Sciences, Department of Physics, University of California San Diego, 9500 Gilman Drive, La Jolla, CA 92093, USA}}
\newcommand{\Belgium}{\affiliation{Sterrenkundig Observatorium, Universiteit Gent, Krijgslaan 281 S9, B-9000 Gent, Belgium}}
\newcommand{\UHeidelberg}{\affiliation{Astronomisches Rechen-Institut, Zentrum für Astronomie der Universität Heidelberg, Mönchhofstr. 12-14, D-69120 Heidelberg, Germany}}
\newcommand{\Bonn}{\affiliation{Argelander-Institut für Astronomie, Universität Bonn, Auf dem Hügel 71, 53121, Bonn, Germany}}
\newcommand{\ANU}{\affiliation{Research School of Astronomy and Astrophysics, Australian National University, Canberra, ACT 2611, Australia}}
\newcommand{\AthreeD}{\affiliation{ARC Centre of Excellence for All Sky Astrophysics in 3 Dimensions (ASTRO 3D), Australia}}
\newcommand{\nrao}{\affiliation{National Radio Astronomy Observatory, 520 Edgemont Road, Charlottesville, VA 22903, USA}}
\newcommand{\ITA}{\affiliation{Institut f\"{u}r Theoretische Astrophysik, Zentrum f\"{u}r Astronomie der Universit\"{a}t Heidelberg,\\ Albert-Ueberle-Strasse 2, 69120 Heidelberg, Germany}}
\newcommand{\COOL}{\affiliation{Cosmic Origins Of Life (COOL) Research DAO, coolresearch.io}}
\newcommand{\IWR}{\affiliation{Universit\"{a}t Heidelberg, Interdisziplin\"{a}res Zentrum f\"{u}r Wissenschaftliches Rechnen, Im Neuenheimer Feld 205, D-69120 Heidelberg, Germany}}
\newcommand{\MPE}{\affiliation{Max-Planck-Institut f\"ur Extraterrestrische Physik (MPE), Giessenbachstr. 1, D-85748 Garching, Germany}}
\newcommand{\ICRAR}{\affil{International Centre for Radio Astronomy Research, University of Western Australia, 35 Stirling Highway, Crawley, WA 6009, Australia}}
\newcommand{\Oxford}{\affil{Sub-department of Astrophysics, Department of Physics, University of Oxford, Keble Road, Oxford OX1 3RH, UK}}
\newcommand{\STScIESA}{\affiliation{AURA for the European Space Agency (ESA), Space Telescope Science Institute, 3700 San Martin Drive, Baltimore, MD 21218, USA}}

\correspondingauthor{Daniel~A.~Dale}
\email{ddale@uwyo.edu}

\author[0000-0002-5782-9093]{Daniel~A.~Dale}
\UWyoming
\author[0000-0003-0946-6176]{M\'ed\'eric~Boquien}
\UAntof
\author[0000-0003-0410-4504]{Ashley~T.~Barnes}
\Bonn
\author[0000-0002-2545-5752]{Francesco Belfiore}
\affiliation{INAF — Arcetri Astrophysical Observatory, Largo E. Fermi 5, I-50125, Florence, Italy}
\author[0000-0003-0166-9745]{F. Bigiel}
\Bonn
\author[0000-0001-5301-1326]{Yixian Cao}
\MPE
\author[0000-0003-0085-4623]{Rupali~Chandar}
\UToledo
\author[0000-0002-5235-5589]{J\'er\'emy~Chastenet}
\Belgium
\author[0000-0002-5635-5180]{M\'elanie Chevance}
\ITA
\COOL
\author[0000-0003-1943-723X]{Sinan~Deger}
\Caltech
\author[0000-0002-4755-118X]{Oleg~V.~Egorov}
\UHeidelberg
\author[0000-0002-3247-5321]{Kathryn~Grasha}
\ANU
\AthreeD
\author[0000-0002-9768-0246]{Brent Groves}
\ICRAR
\author[0000-0002-8806-6308]{Hamid~Hassani}
\UAlberta
\author[0000-0001-7448-1749]{Kiana~F.~Henny}
\UWyoming
\author[0000-0002-0560-3172]{Ralf S.\ Klessen}
\ITA
\IWR
\author[0000-0001-6551-3091]{Kathryn Kreckel}
\UHeidelberg
\author[0000-0002-8804-0212]{J.~M.~Diederik~Kruijssen}
\COOL
\author[0000-0003-3917-6460]{Kirsten~L.~Larson}
\STScIESA
\author[0000-0003-0946-6176]{Janice~C.~Lee}
\NOIRLab
\author[0000-0002-2545-1700]{Adam~K.~Leroy}
\OSU
\author[0000-0001-9773-7479]{Daizhong Liu}
\MPE
\author[0000-0001-7089-7325]{Eric J.\,Murphy}
\nrao
\author[0000-0002-5204-2259]{Erik~Rosolowsky}
\UAlberta
\author[0000-0002-4378-8534]{Karin~Sandstrom}
\UCSD
\author[0000-0002-3933-7677]{Eva~Schinnerer}
\MPIA
\author[0000-0002-9183-8102]{Jessica Sutter}
\UCSD
\author[0000-0002-8528-7340]{David~A.~Thilker}
\JHU
\author[0000-0002-7365-5791]{Elizabeth~J.~Watkins}
\UHeidelberg
\author{Bradley~C.~Whitmore}
\STScI
\author[0000-0002-0012-2142]{Thomas~G.~Williams}
\Oxford
\MPIA




\begin{abstract}
We present a comparison of theoretical predictions of dust continuum and polycyclic aromatic hydrocarbon (PAH) emission with new \jwst\ observations in three nearby galaxies: NGC~628, NGC~1365, and NGC~7496.  Our analysis focuses on a total of 1063 compact stellar clusters and 2654 stellar associations previously characterized by \hst\ in the three galaxies.  We find that the distributions and trends in the observed PAH-focused infrared colors generally agree with theoretical expectations, and that the bulk of the observations is more aligned with models of larger, ionized PAHs.  These \jwst\ data usher in a new era of probing interstellar dust and studying how the intense radiation fields near stellar clusters and associations play a role in shaping the physical properties of PAHs.
\end{abstract}




\section{Introduction} \label{sec:intro}
 Polycyclic aromatic hydrocarbon (PAHs) contribute as much as $20$\% of a star-forming galaxy's total infrared emission \citep{smith2007} and produce bright emission features in the mid-infrared \citep[e.g.,][]{tielens2008} that have been used as star formation tracers across cosmic time \citep[e.g.,][]{riechers2014}.  However, PAH emission exhibits large variability within and between galaxies, and it is important to determine the factors that drive PAH properties and strength.  Previous studies of PAH emission features in galaxies have shown that the strength of PAH emission is sensitive to the metal abundance of the interstellar medium, the molecular gas surface density, the hardness and strength of the interstellar radiation field (or related metrics such as the star formation rate, dust temperature, active galactic nucleus X-ray luminosity, etc.), or some combination thereof \citep[e.g.,][]{engelbracht2005,madden2006,wu2006,draine2007,engelbracht2008,dale2009,sandstrom2010,wu2010,remyruyer2015,jensen2017,chastenet2019,aniano2020,galliano2021,wolfire2022}.  These studies that were completed using previous generations of space-based infrared facilities targeted bright peaks of infrared emission or averaged over large swaths of lower surface brightness regions.  The advent of the more sensitive instruments aboard \jwst\ enable characterization of the PAH emission at high angular resolution across entire galaxy disks, including the 3.3~\m\ PAH feature which is essential to constraining the PAH size distribution but heretofore rarely observed for statistically significant samples \citep[][and references therein]{lai2020}.  Similarly, the relative strengths of PAH emission features, including those at 3.3, 7.7, and 11.2~\m, have long been thought to be critically dependent on the levels of PAH ionization \citep[e.g.,][]{defrees1989,allamondola1999}.

With JWST and HST we present a novel study of the relationship between the strength of three key PAH mid-infrared features on the physical scales of star clusters and associations, and investigate trends with the ages of those stellar populations.  Intense ultraviolet radiation from star clusters dominated by young OB stars will, for example, ionize or even destroy PAHs, which in turn will impact the local reddening curve, the efficiency of the heating of the neutral interstellar medium, chemical reaction rates, etc. \citep{hollenbach1997,draine2007,tielens2008,wolfire2022}.  We compare the theoretical predictions for PAH emission features with photometric observations from \jwst\ for the Physics at High Angular resolution in Nearby GalaxieS (PHANGS) program on nearby galaxies \citep{leroy2021,emsellem2022,lee2022,LEE22JWST}, using three filters with bandpasses that capture PAH emission features centered at 3.3, 7.7, and 11.2~\m.  We specifically compare our observations with the predictions laid out in \cite{draine2021} \citep[see also][]{rigopoulou2021} for a range of PAH ionization levels and size distributions in addition to a variety of interstellar radiation field intensities and ages of the stellar populations that drive the dust heating.


\section{Sample and Data} \label{sec:sample_data}
The galaxies and observations analyzed here are drawn from the PHANGS project.  The first wave of PHANGS--\jwst\ data included a suite of NIRCam \citep{rieke2005} and MIRI \citep{rieke2015} imaging for three star-forming galaxies.  NGC~628 is a grand-design spiral galaxy at 9.84~Mpc, and both NGC~1365 (19.57~Mpc) and NGC~7496 (18.72~Mpc) are barred spirals with Seyfert nuclei (distances are from \citealt{shaya2017,kourkchi2017,anand2021,anand2021b}).  The NIRCam mosaics use the F200M, F300M, F335M, and F360M filters and the MIRI mosaics use the F770W, F1000W, F1130W, and F2100W filters.  The technical description and post-processing of the NIRCam and MIRI imaging for the PHANGS-\jwst\ program is described in \cite{LEE22JWST}.  

The catalog of stellar clusters utilized here is from the PHANGS--\hst\ program \citep{turner2021,lee2022,thilker2022}; we restrict our analysis to the Class~1 and 2 clusters \citep{whitmore2021,deger2022} which are either compact and centrally concentrated (Class~1) or compact and slightly asymmetric (Class~2).  In lieu of using the catalog of asymmetric and multi-peaked Class~3 stellar associations, which is comparatively incomplete since the PHANGS--\hst\ pipeline was optimized for detecting single-peaked compact clusters, we utilize the PHANGS--\hst\ catalog of stellar associations.  These stellar associations, which are less likely to be gravitationally bound than Class~1 and 2 compact clusters \citep{whitmore2021}, are derived from a watershed analysis that is based on $V$-band point-source detections and a 32~pc full-width-half-maximum Gaussian smoothing, as presented in \citep{LEE22JWST} and Larson et al.\ (in prep). The stellar masses and ages for the clusters and associations are based on spectral energy distribution fits to the five-band ultraviolet/optical PHANGS--\hst\ datasets and span $\sim 10^{2.8-7.0}~M_\odot$ and 1~Myr to 13~Gyr \citep[see also][]{turner2021}.  

\begin{figure*}
 \plotone{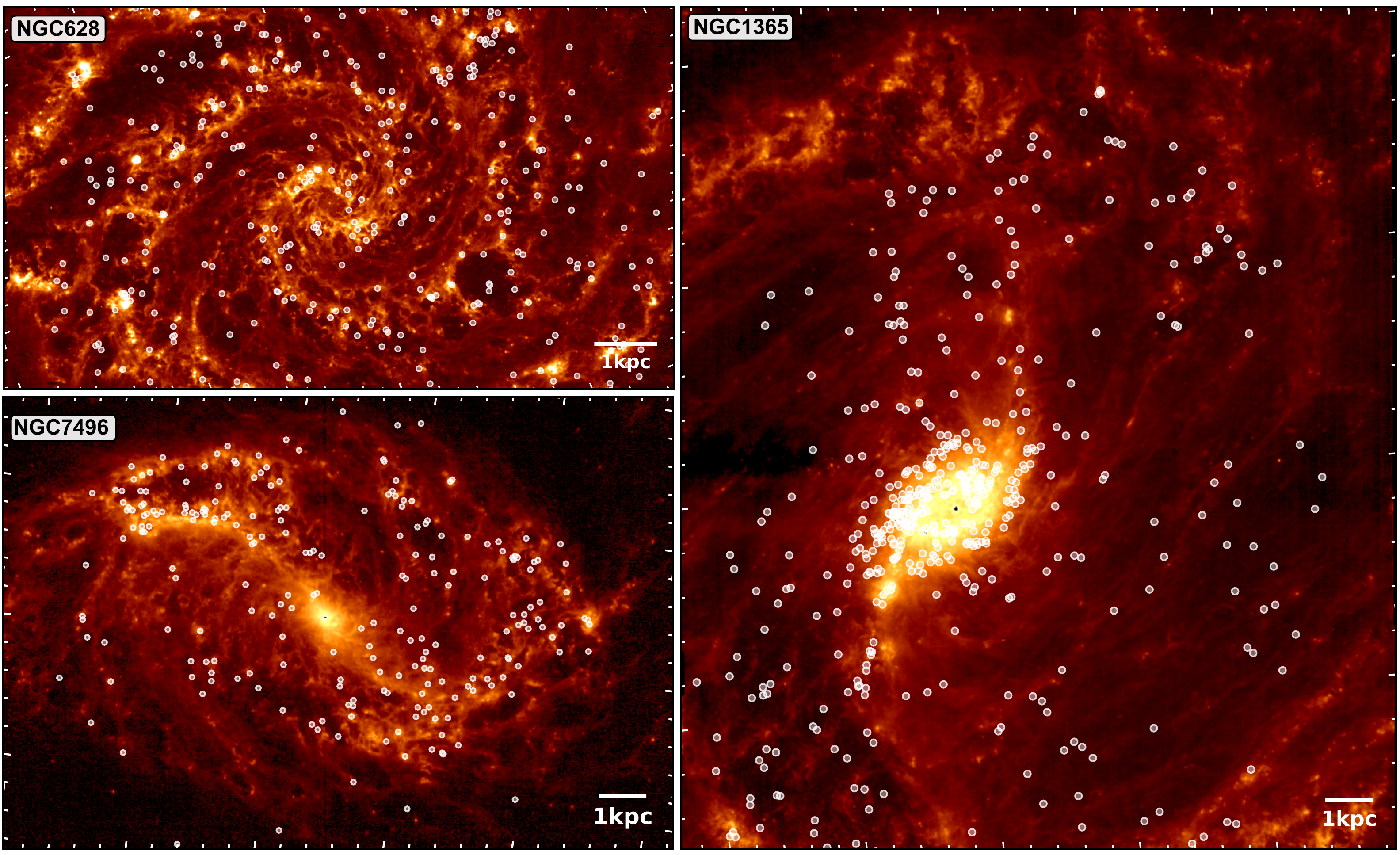}
 \caption{F770W imaging along with the locations of the stellar clusters.  A 1~kpc scale bar is included within each panel.}
 \label{fig:mosaic}
\end{figure*}

There are a total of $330 + 501 + 232 = 1063$ compact clusters and $1539+689+426=2654$ stellar associations in the publicly-available catalogs that overlap with the PHANGS--\jwst\ footprints for NGC~628, NGC~1365, and NGC~7496, respectively.  As can be seen in Figure~\ref{fig:mosaic}, these three galaxies are rich with mid-infrared emission throughout their disks; 95\% of the compact clusters and 93\% of the stellar associations show mid-infrared emission coincident with their locations in at least one of the three bandpasses. 

We remove clusters and associations for which their mid-infrared photometry is potentially impacted by the saturation effects associated with the active galactic nuclei for NGC~1365 and NGC~7496.  Following \cite{HASSANI22JWST}, we reject sources that overlap with regions where the amplitude of the modeled PSFs, centered on the locations of the AGN, are $>0.1\%$ of the PSF maxima.  These masked central regions span 0.32~kpc$^2$ and 0.33~kpc$^2$, respectively, for NGC~1365 and NGC~7496.  This step removes 6 clusters and 21 associations for NGC~1365 near the AGN, and 12 clusters and 33 associations for NGC~7496.  If we further require a stellar mass of $\log (M_*/M_\odot) > 3.3$ to align with our targeted median cluster mass sensitivity for \jwst\ observations \citep[][Larson et al.\ in prep]{lee2022}, then the cluster sample size drops another 5\% and the association sample drops another 17\%.


\section{Analysis} \label{sec:analysis}
As described in \cite{LEROY22JWST}, we first make small background level corrections to the imaging using galaxy-free regions in the outer portions of each image.  Since our analysis relies on a combination of the F300M, F335M, F360M, F770W, and F1130W images, all the imaging was convolved with smoothing kernels following \cite{aniano11} to achieve the effective 0\farcs37 angular resolution of the F1130W data.  Continuum-subtracted 3.3~\m\ PAH maps (F335M$_{\rm PAH}$) are constructed using the method outlined in \cite{SANDSTROM22JWST} that leverages the F300M and F360M imaging to infer the underlying continua.  No continuum subtraction is carried out for the F770W or F1130W imaging since the stellar contributions at these wavelengths are minor and the dust emission appearing in these two bands is dominated by PAH features \citep{smith2007,EGOROV22JWST,HASSANI22JWST}.  Future work will explore the impact of using, for example, the F1000W imaging as a proxy for the wavelength-adjacent continuum for the F770W and F1130W bands.

Aperture photometry is carried out for each compact stellar cluster using an aperture radius of 0\farcs3.  The general trends discussed in \S~\ref{sec:results} are unaffected by modifications to our choice for aperture radius.  The photometry for each stellar association is a simple sum of the fluxes for each pixel within the association's defined polygon, the edges for which are based on the surface `brightness' of the smoothed tracer star maps (see Larson et al.\ 2023, in prep, for details).  The choice of the 32~pc-scale stellar association watershed maps matches well with the apertures utilized for the compact stellar clusters---for our galaxy sample's distances, 0\farcs3 radii correspond to 14--28~pc.  No local ``background'' subtraction is applied to the cluster or association photometry.

\begin{figure}
 \plotone{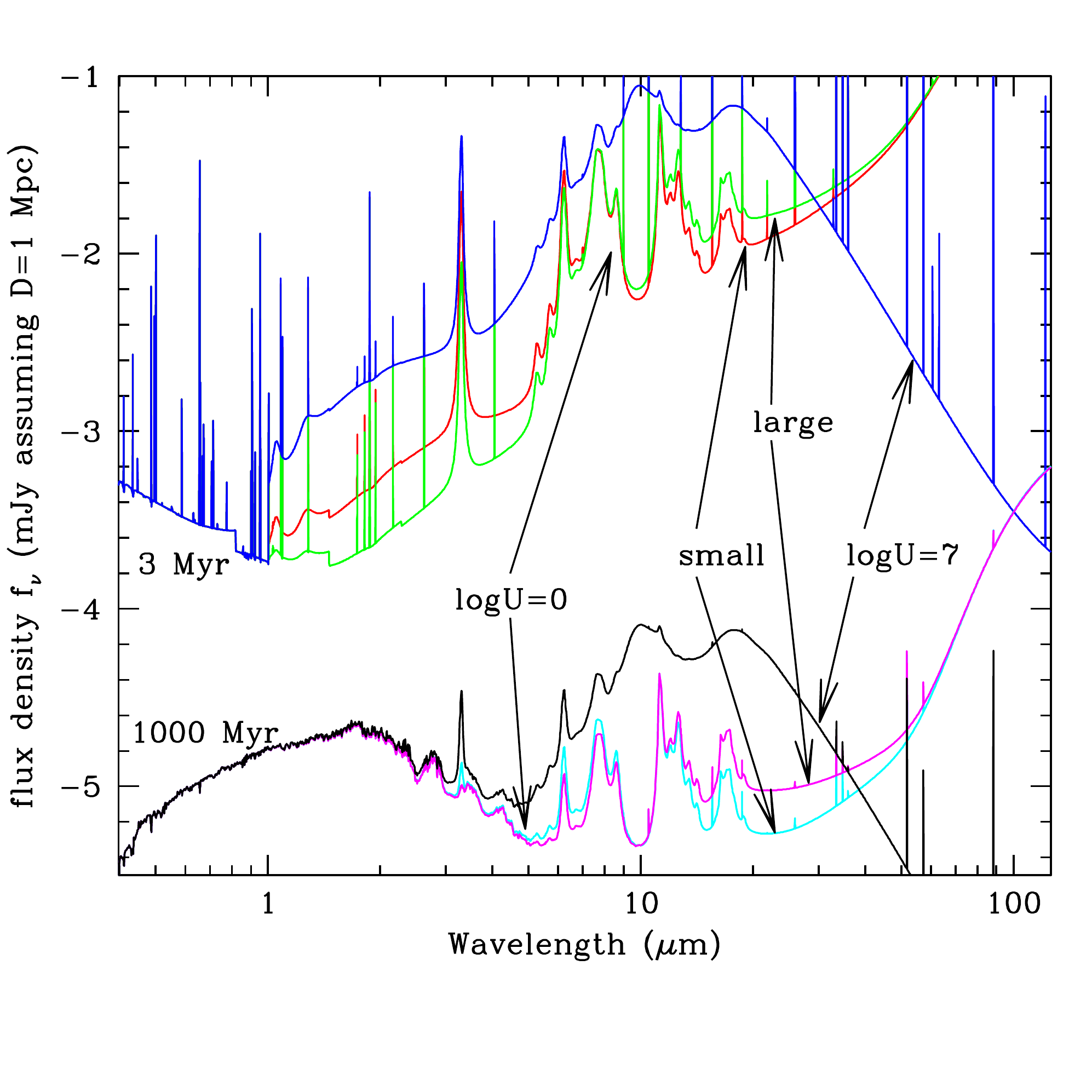}
 \caption{Example synthetic spectra drawn from the dust models of \cite{draine2021} and generated using the CIGALE software (see \S~\ref{sec:analysis} for details).  The spectra are shown for maximum differences in the chosen grid for stellar age, PAH size, and interstellar radiation field intensity.  The changes in the spectra as a function of PAH ionization are less pronounced and are thus not displayed.  The six different colors represent the six different example spectra, with their physical properties indicated by arrows and their descriptors.}
 \label{fig:seds}
\end{figure}

\cite{draine2021} present various PAH band ratio diagrams for comparison with observations (e.g., see their Figures~16--21).  Since the models of \cite{draine2021} do not include contributions from stars, nebular lines, or other sources such as active galactic nuclei, in their analysis they employ a ``clipping'' method to focus solely on the PAH emission features.  Because our observed photometry in the \jwst\ bandpasses inevitably include contributions from such other non-PAH sources of emission, we make direct comparisons with the models by extracting synthetic fluxes for each bandpass using the CIGALE software \citep{boquien2019} and the \cite{draine2021} models as the input simulated spectra.  To simulate a variety of conditions we construct a suite of simulated spectra that span a range of PAH ionizations ($\texttt{ion}=0,1,2$; see \citealt{draine2021} Figure~9b), PAH size distributions ($\texttt{size}=0,1,2$ corresponding to $a_{01}=3,4,5~$\AA; see \citealt{draine2021} Figure~9a), interstellar radiation field intensities $U$ ($\texttt{logU}=0-7$), and age of the main stellar population ($\texttt{age}=3,10,100,1000$~Myr).  We assume solar metallicity \cite{bruzual2003} stellar populations with a standard \cite{chabrier2003} initial mass function.  Six of these synthetic spectra, chosen to demonstrate the dynamic range available in the models, are shown in Figure~\ref{fig:seds}.  The other parameters in CIGALE are fixed since varying them does not significantly change our results or interpretation, e.g., opting for an exponential star formation history with a late burst, solar metallicity, a Lyman continuum photon escape fraction of 0, etc.  Finally, we have applied to our synthetic data the prescription outlined in \cite{SANDSTROM22JWST} for removing the underlying continuum to the 3.3\m\ PAH feature emission, to be consistent with our treatment of the observations.

\section{Results and Discussion} \label{sec:results}
\begin{figure*}
 \plotone{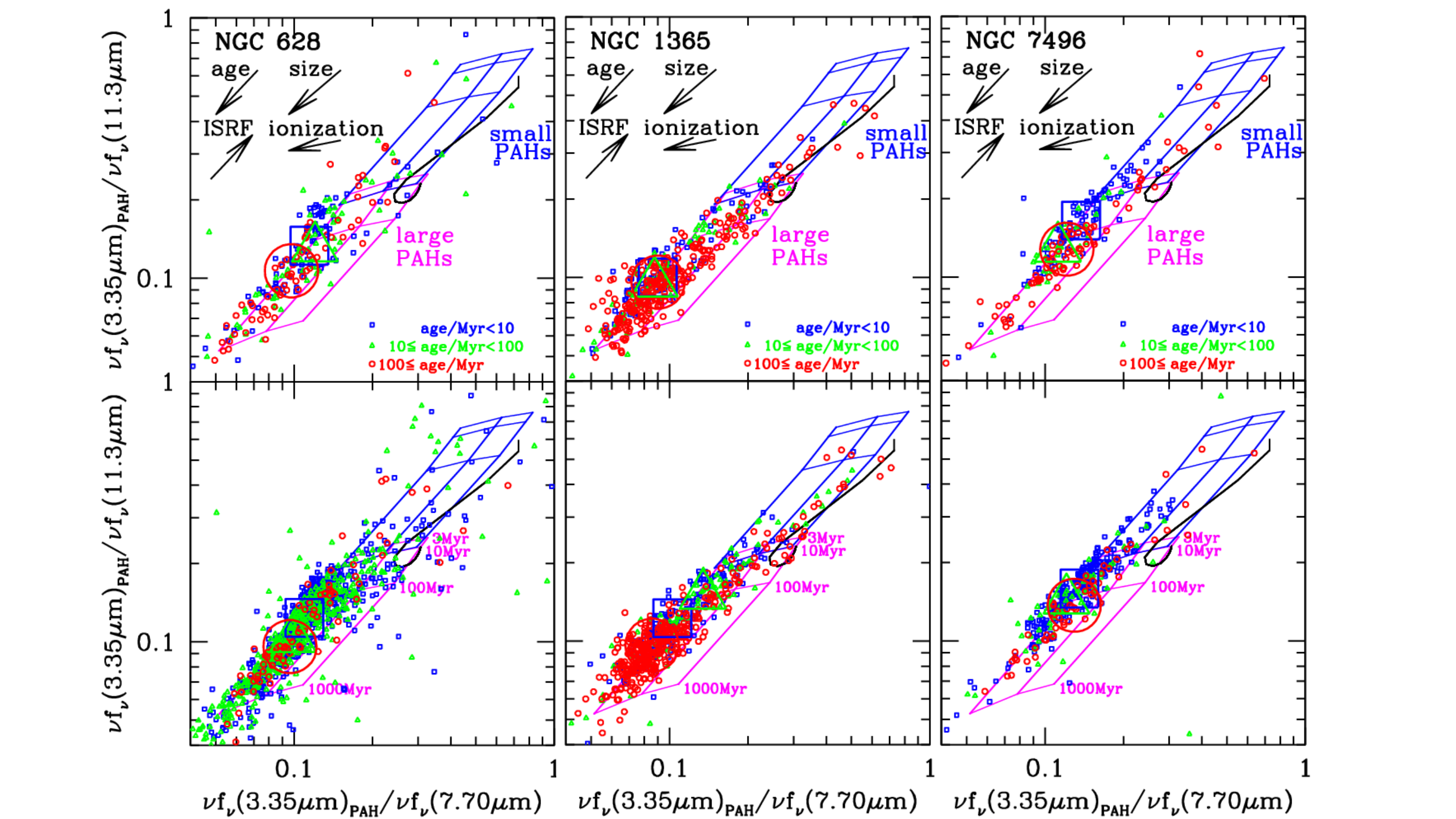}
 \caption{PAH bandpass flux ratios for the compact clusters (top) and stellar associations (bottom).  The open blue squares, green triangles, and red circles are for compact stellar clusters, colored according to their ages: 0--10~Myr, 10--100~Myr, and $>$100~Myr, respectively.  Large symbols indicate medians.  Overlaid tracks are presented for select subsets of the simulated ratios, and the general trends for these tracks are indicated with the inset arrows/descriptions (see text for details).  The magenta and blue tracks are for small ($a_{01}=3~$\AA) and large ($a_{01}=5~$\AA) modeled PAH distributions, respectively, with both assuming \texttt{logU}=0, \texttt{ion}=0,1.2, and \texttt{age}=10~Myr.  The black hook-shaped track spans $\log U$=0,7 and fixes \texttt{ion}=0, \texttt{size}=2, and \texttt{age}=10~Myr.}
 \label{fig:three_panel}
\end{figure*}

Figure~\ref{fig:three_panel} provides the F335M$_{\rm PAH}$/F1130W and F335M$_{\rm PAH}$/F770W band ratios for the three galaxies, where we use the shorthand notation ``F335M$_{\rm PAH}$'' to represent $\nu f_\nu$(F335M) for the PAH feature flux extracted for the F335M filter, etc.  The data are color-coded according to stellar age.  The two observed ratios each span about 1.4~dex, and they scale approximately linearly with each other; the numerators in the ratios are the same, and the denominators are comparable---thus the approximately one-to-one correlation seen in Figure~\ref{fig:three_panel}.  We provide in Table~\ref{tab:medians} the population medians and their semi-interquartile spreads in log base 10, along with the population means and their standard deviations.  In addition, we provide the results of $t$-tests to quantify how significantly different are the population means.  For the compact clusters in NGC~628 and NGC~7496 the medians and means for the F335M$_{\rm PAH}$/F1130W and F335M$_{\rm PAH}$/F770W ratios decrease with increasing stellar age as suggested by the synthetic model extractions described immediately below, particularly when comparing the $<$10~Myr and $>$100~Myr populations.  For the stellar associations the overall distributions are consistent with what is observed for the compact clusters, and the older stellar populations show noticeably lower F335M$_{\rm PAH}$/F1130W and F335M$_{\rm PAH}$/F770W ratios in NGC~628 and NGC~1365.  However, the $t$-test results imply that these differences are only statistically significant for the associations in NGC~628 and NGC~1365 and marginally significant for the clusters in NGC~628.  Finally, we note that the scatter in the distributions increases with stellar age; the average semi-interquartile ranges for the young, intermediate, and old clusters and associations are 0.097, 0.122, and 0.124~dex and the average standard deviations are 0.216, 0.256, and 0.307~dex, respectively, for the three age populations.

\begin{deluxetable*}{llcccl}
\tabletypesize{\scriptsize}
\tablecaption{Statistics for Figure~\ref{fig:three_panel} \label{tab:medians}}
\tablewidth{0pc}
\tablehead{
\colhead{Galaxy} &
\colhead{Population} &
\colhead{age/Myr$<$10} &
\colhead{10$\leq$age/Myr$<$100} &
\colhead{100$\leq$age/Myr} &
\colhead{probability}
}
\startdata
&&$^\dagger$med$_x$,med$_y$~~SIQR$_x$,SIQR$_y$ & med$_x$,med$_y$~~SIQR$_x$,SIQR$_y$ & med$_x$,med$_y$~~SIQR$_x$,SIQR$_y$ & Kolm-Smir\\
\hline
NGC0628 & clusters     & $-$0.94,$-$0.88 ~~~ 0.08,0.13 & $-$0.92,$-$0.88 ~~~~ 0.12,0.15 & $-$1.01,$-$0.97 ~~~~ 0.15,0.17 & 7$\cdot 10^{-3}$\\
NGC0628 & associations & $-$0.96,$-$0.91 ~~~ 0.09,0.12 & $-$1.00,$-$0.99 ~~~~ 0.13,0.15 & $-$1.02,$-$1.02 ~~~~ 0.14,0.16 & 2$\cdot 10^{-4}$\\
NGC1365 & clusters     & $-$1.05,$-$1.00 ~~~ 0.08,0.08 & $-$1.06,$-$1.02 ~~~~ 0.12,0.14 & $-$1.04,$-$1.02 ~~~~ 0.11,0.10 & 1$\cdot 10^{-3}$\\
NGC1365 & associations & $-$0.99,$-$0.91 ~~~ 0.11,0.12 & $-$0.87,$-$0.82 ~~~~ 0.14,0.13 & $-$1.07,$-$1.02 ~~~~ 0.09,0.07 & 6$\cdot 10^{-8}$\\
NGC7496 & clusters     & $-$0.86,$-$0.78 ~~~ 0.09,0.10 & $-$0.95,$-$0.89 ~~~~ 0.09,0.09 & $-$0.92,$-$0.89 ~~~~ 0.15,0.14 & 6$\cdot 10^{-4}$\\
NGC7496 & associations & $-$0.87,$-$0.80 ~~~ 0.08,0.09 & $-$0.92,$-$0.84 ~~~~ 0.12,0.09 & $-$0.89,$-$0.86 ~~~~ 0.10,0.12 & 6$\cdot 10^{-3}$\\
\hline
&& $^\ddagger$mean$_x$,mean$_y$ ~~ $\sigma_x$,$\sigma_y$ & mean$_x$,mean$_y$ ~~ $\sigma_x$,$\sigma_y$ & mean$_x$,mean$_y$ ~~ $\sigma_x$,$\sigma_y$ & $t$-test\\
\hline
NGC0628 & clusters     & $-$0.94,$-$0.89 ~~~ 0.22,0.24 & $-$0.88,$-$0.86 ~~~~ 0.35,0.27 & $-$1.01,$-$0.97 ~~~~ 0.36,0.37 & 0.16,0.10\\
NGC0628 & associations & $-$0.96,$-$0.94 ~~~ 0.26,0.27 & $-$1.00,$-$1.00 ~~~~ 0.29,0.30 & $-$1.03,$-$1.02 ~~~~ 0.44,0.45 & 0.07,0.04\\
NGC1365 & clusters     & $-$1.02,$-$0.98 ~~~ 0.20,0.20 & $-$1.02,$-$1.00 ~~~~ 0.27,0.28 & $-$0.99,$-$1.00 ~~~~ 0.26,0.24 & 0.11,0.70\\
NGC1365 & associations & $-$0.94,$-$0.90 ~~~ 0.22,0.22 & $-$0.87,$-$0.85 ~~~~ 0.27,0.25 & $-$1.02,$-$1.00 ~~~~ 0.26,0.26 & 0.03,0.0009\\
NGC7496 & clusters     & $-$0.89,$-$0.81 ~~~ 0.20,0.22 & $-$0.96,$-$0.91 ~~~~ 0.18,0.20 & $-$0.85,$-$0.86 ~~~~ 0.32,0.26 & 0.33,0.25\\
NGC7496 & associations & $-$0.88,$-$0.81 ~~~ 0.17,0.18 & $-$0.89,$-$0.86 ~~~~ 0.20,0.22 & $-$0.84,$-$0.82 ~~~~ 0.23,0.23 & 0.26,0.63\\
\hline
\enddata
\tablenotetext{}{$^\dagger$med$_x$ and med$_y$ refer to the medians of log$_{10}$(F335M$_{\rm PAH}$/F770W) and log$_{10}$(F335M$_{\rm PAH}$/F1130W), respectively. The semi-interquartile ranges are  half the 25\%--75\% range found after sorting the flux ratios.  The SIQRs provided here are in units of dex since they are computed using the (base 10) logarithms of the flux ratios.  The Kolmogorov-Smirnov probabilities characterize how significantly the $<$10~Myr and $>$100~Myr two-dimensional distributions differ, with probabilities less than $\sim 0.05$ indicating a significant difference \citep{press1993}.}
\tablenotetext{}{$^\ddagger$mean$_x$ and mean$_y$ refer to the means of log$_{10}$(F335M$_{\rm PAH}$/F770W) and log$_{10}$(F335M$_{\rm PAH}$/F1130W), respectively; $\sigma$ indicates the standard deviation.  $t$-test probabilities characterize how significantly the $<$10~Myr and $>$100~Myr (one-dimensional log$x$,log$y$) means differ, with probabilities less than $\sim 0.05$ indicating significantly different values \citep{press1993}.}
\end{deluxetable*}

We also include in Figure~\ref{fig:three_panel} the grid of synthetic points from the models of \cite{draine2021} described in \S~\ref{sec:analysis}.  The magenta and blue grids in Figure~\ref{fig:three_panel} demonstrate how the synthetic points depend on stellar cluster age, PAH size distribution, and PAH ionization (after fixing $U=1$).  The black track portrays how the synthetic data vary with the intensity of the interstellar radiation field for $\log U =0-7$ (after fixing $\texttt{ion}=0$, $\texttt{size}=2$, and $\texttt{age}=10$~Myr).  The large arrows and corresponding descriptors indicate how the synthetic tracks change with each modeled parameter; these synthetic trends are consistent with those portrayed in Figure~21d of \cite{draine2021}.  The overall distributions of the observed ratios in Figure~\ref{fig:three_panel} track well with the displayed grid of synthetic models, and the majority of the observations are consistent with the PAHs having elevated ionization levels and large size distributions.  In other words, the data are mostly populating the magenta grid for larger PAHs (and not the blue grid for smaller PAHs), between the magenta model tracks for 100~Myr and 1000~Myr, and the data are more closely aligned with the lefthand edge of the magenta grid that indicates more highly ionized PAHs.  The implication is that the proximity of stellar clusters and associations has enhanced the photo-ejection of electrons from the PAHs.  

Our evidence for higher levels of PAH ionization is conceptually consistent with the work of other PHANGS--\jwst\ efforts appearing in this issue.  \cite{EGOROV22JWST} analyze the F1130W/F770W parameter,  widely considered to be a tracer of the ratio of neutral to ionized PAHs \citep{draine2001, maragkoudakis2020}, in HII regions in NGC~628, NGC~1365, NGC~7496, and IC~5332 and find lower values of F1130W/F770W in regions with higher \SIII/\SII\ ratios, indicating harder radiation fields.  \cite{CHASTENET22JWST} probe PAH band ratios across the disks of the four galaxies and find evidence of hotter, highly ionized PAHS in the vicinity of \HII\ regions which are defined by the radiation from young stars.  Finally, \cite{SANDSTROM22JWST} create maps of PAH band ratios in the same three galaxies considered in this work and find fairly flat F335M$_{\rm PAH}$/F1130W radial profiles, with amplitudes similar to what we find (0.07--0.24 in $\nu f_\nu$ units).  A follow-up control study using a larger sample of PHANGS galaxies will compare the PAH ionization levels inferred in this work near stellar clusters and associations with those of PAHs that are spread throughout the diffuse interstellar medium.

Close scrutiny of Figure~\ref{fig:three_panel} suggests subtle differences between galaxies, such as the sources in NGC~7496 being on average higher up in the magenta grid which is consistent with more intense radiation fields (and/or young stellar ages).  It is unlikely that the AGN in NGC~7496 is responsible for this difference, as we have removed the central clusters and associations from our analysis (see \S~\ref{sec:sample_data}) and our stellar sources in NGC~7496 are dispersed throughout its disk and spiral arms (Figure~\ref{fig:mosaic}).  In addition, NGC~628 shows larger scatter in the data than for NGC~1365 and NGC~7496, with standard deviations in the abscissa and ordinate values larger by a factor of 1.23, 1.30, and 1.56 for the $<$10~Myr, 10--100~Myr, and $>$100~Myr populations, respectively.  The factor of two larger distances for NGC~1365 and NGC~7496 compared to the distance for NGC~628 result in more spatial averaging over the 0\farcs6 diameter aperture and thus the smaller scatter in the data.


\section{Conclusions} \label{sec:conclusions}
We have leveraged new \jwst\ near- and mid-infrared imaging of NGC~628, NGC~1365, and NGC~7496 to study their PAH emission.  With the combined data from the three galaxies we are able to analyze the PAH emission over localized ($<32$~pc scale) regions centered on 1063 compact stellar clusters and 2654 stellar associations previously analyzed with HST ultraviolet/optical imaging.   This study represents the first look at matched PAH band ratios, at high angular resolution, for large statistical samples of stellar ionizing sources isolated from more diffuse ISM conditions, enabling us to directly link ionizing sources to their impact on PAH properties across representative disk environments.

To enable a comparison with theoretical expectations, we extracted synthetic infrared colors based on models of stellar, PAH, and dust continuum emission.  A full grid of synthetic values is created by modifying four key parameters: stellar age, PAH ionization fraction, PAH size distribution, and interstellar radiation field intensity.  Our results are generally consistent with the predictions from the dust models outlined in \cite{draine2021}, though the PAH band ratios we measure for the youngest stellar clusters and associations ($<10$~Myr) appear between theoretical grid values (assuming large PAH size distributions) for stellar ages between 100~Myr and 1000~Myr.  In addition, we find statistical evidence for the stellar associations in NGC~628 and NGC~1365 having decreasing values of F335M$_{\rm PAH}$/F1130W and F335M$_{\rm PAH}$/F770W with increasing stellar age, consistent with the expectations based on the synthetic extractions.  The slope and the overall distribution for the ensemble of synthetic PAH band ratios is consistent with that for the observed distributions in the three galaxies, and the models with higher levels of PAH ionization and larger size distributions are generally more aligned with the observations.  A higher level of ionization could indicate enhanced processing of PAHs from the radiation fields produced by the stellar clusters.  Finally, there is an age-dependent trend in the scatter, with older stellar populations exhibiting larger dispersions in the PAH band ratios.

Further additional analysis with the full PHANGS--\jwst\ sample is needed to more firmly test theoretical dust models.  The larger PHANGS sample will enable us to explore PAH band ratios as a function of local metallicity and radiation hardness indicators, and thus bring additional insight into the dearth of PAHs for environments of depressed metal abundance and/or elevated radiation hardness.  Furthermore, an analysis that leverages the larger PHANGS-\jwst\ sample will provide a more robust control sample for measuring PAH band ratios in diffuse regions as a control sample against which we will compare our measurements near stellar clusters and associations.


\section{acknowledgments}
We thank the referee for their thoughtful review.  This work is based on observations made with the NASA/ESA/CSA JWST and Hubble Space Telescopes.  The data were obtained from the Mikulski Archive for Space Telescopes at the Space Telescope Science Institute, which is operated by the Association of Universities for Research in Astronomy, Inc., under NASA contract NAS 5-03127 for JWST and NASA contract NAS 5-26555 for HST.  The JWST observations are associated with Program~2107, and those from HST with Program~15454.  The specific observations analyzed can be accessed via \dataset[10.17909/9bdf-jn24]{http://dx.doi.org/10.17909/9bdf-jn24}.  The HST images and catalogs can be access via \dataset[10.17909/t9-r08f-dq31]{https://dx.doi.org/10.17909/t9-r08f-dq31} and \dataset[10.17909/jray-9798]{https://dx.doi.org/10.17909/jray-9798}, respectively.
FB would like to acknowledge funding from the European Research Council (ERC) under the European Union’s Horizon 2020 research and innovation programme (grant agreement No.726384/Empire).
MB acknowledges support from FONDECYT regular grant 1211000 and by the ANID BASAL project FB210003.
JC acknowledges support from ERC starting grant \#851622 DustOrigin.
MC gratefully acknowledges funding from the Deutsche Forschungsgemeinschaft (DFG) through an Emmy Noether Research Group (grant number CH2137/1-1).
KG is supported by the Australian Research Council through the Discovery Early Career Researcher Award (DECRA) Fellowship DE220100766 funded by the Australian Government. 
KG is supported by the Australian Research Council Centre of Excellence for All Sky Astrophysics in 3 Dimensions (ASTRO~3D), through project number CE170100013. 
JMDK gratefully acknowledges funding from the European Research Council (ERC) under the European Union's Horizon 2020 research and innovation programme via the ERC Starting Grant MUSTANG (grant agreement number 714907). COOL Research DAO is a Decentralized Autonomous Organization supporting research in astrophysics aimed at uncovering our cosmic origins.
KK, OE gratefully acknowledge funding from the Deutsche Forschungsgemeinschaft (DFG, German Research Foundation) in the form of an Emmy Noether Research Group (grant number KR4598/2-1, PI Kreckel).
RSK acknowledges financial support from the European Research Council via the ERC Synergy Grant ``ECOGAL'' (project ID 855130), from the Deutsche Forschungsgemeinschaft (DFG) via the Collaborative Research Center ``The Milky Way System'' (SFB 881---funding ID 138713538---subprojects A1, B1, B2 and B8) and from the Heidelberg Cluster of Excellence (EXC 2181---390900948) ``STRUCTURES'', funded by the German Excellence Strategy. RSK also thanks the German Ministry for Economic Affairs and Climate Action for funding in the project ``MAINN'' (funding ID 50OO2206). 
AKL gratefully acknowledges support by grants 1653300 and 2205628 from the National Science Foundation, by award JWST-GO-02107.009-A, and by a Humboldt Research Award from the Alexander von Humboldt Foundation.
ER acknowledges the support of the Natural Sciences and Engineering Research Council of Canada (NSERC), funding reference number RGPIN-2022-03499.
EJW acknowledges the funding provided by the Deutsche Forschungsgemeinschaft (DFG, German Research Foundation) -- Project-ID 138713538 -- SFB 881 (``The Milky Way System'', subproject P1). 
TGW acknowledges funding from the European Research Council (ERC) under the European Union’s Horizon 2020 research and innovation programme (grant agreement No. 694343).


\bibliography{main.bib}

\begin{thebibliography}{}
\expandafter\ifx\csname natexlab\endcsname\relax\def\natexlab#1{#1}\fi
\providecommand{\url}[1]{\href{#1}{#1}}

\bibitem[{{Allamandola} {et~al.}(1999){Allamandola}, {Hudgins}, \&
  {Sandford}}]{allamondola1999}
{Allamandola}, L.~J., {Hudgins}, D.~M., \& {Sandford}, S.~A. 1999, \apjl, 511,
  L115

\bibitem[{{Anand} {et~al.}(2021{\natexlab{a}}){Anand}, {Lee}, {Van Dyk},
  {Leroy}, {Rosolowsky}, {Schinnerer}, {Larson}, {Kourkchi}, {Kreckel},
  {Scheuermann}, {Rizzi}, {Thilker}, {Tully}, {Bigiel}, {Blanc}, {Boquien},
  {Chandar}, {Dale}, {Emsellem}, {Deger}, {Glover}, {Grasha}, {Groves}, {S.
  Klessen}, {Kruijssen}, {Querejeta}, {S{\'a}nchez-Bl{\'a}zquez}, {Schruba},
  {Turner}, {Ubeda}, {Williams}, \& {Whitmore}}]{anand2021}
{Anand}, G.~S., {Lee}, J.~C., {Van Dyk}, S.~D., {et~al.} 2021{\natexlab{a}},
  \mnras, 501, 3621

\bibitem[{{Anand} {et~al.}(2021{\natexlab{b}}){Anand}, {Rizzi}, {Tully},
  {Shaya}, {Karachentsev}, {Makarov}, {Makarova}, {Wu}, {Dolphin}, \&
  {Kourkchi}}]{anand2021b}
{Anand}, G.~S., {Rizzi}, L., {Tully}, R.~B., {et~al.} 2021{\natexlab{b}}, \aj,
  162, 80

\bibitem[{{Aniano} {et~al.}(2011){Aniano}, {Draine}, {Gordon}, \&
  {Sandstrom}}]{aniano11}
{Aniano}, G., {Draine}, B.~T., {Gordon}, K.~D., \& {Sandstrom}, K. 2011, \pasp,
  123, 1218

\bibitem[{{Aniano} {et~al.}(2020){Aniano}, {Draine}, {Hunt}, {Sandstrom},
  {Calzetti}, {Kennicutt}, {Dale}, {Galametz}, {Gordon}, {Leroy}, {Smith},
  {Roussel}, {Sauvage}, {Walter}, {Armus}, {Bolatto}, {Boquien}, {Crocker}, {De
  Looze}, {Donovan Meyer}, {Helou}, {Hinz}, {Johnson}, {Koda}, {Miller},
  {Montiel}, {Murphy}, {Rela{\~n}o}, {Rix}, {Schinnerer}, {Skibba}, {Wolfire},
  \& {Engelbracht}}]{aniano2020}
{Aniano}, G., {Draine}, B.~T., {Hunt}, L.~K., {et~al.} 2020, \apj, 889, 150

\bibitem[{{Boquien} {et~al.}(2019){Boquien}, {Burgarella}, {Roehlly}, {Buat},
  {Ciesla}, {Corre}, {Inoue}, \& {Salas}}]{boquien2019}
{Boquien}, M., {Burgarella}, D., {Roehlly}, Y., {et~al.} 2019, \aap, 622, A103

\bibitem[{{Bruzual} \& {Charlot}(2003)}]{bruzual2003}
{Bruzual}, G., \& {Charlot}, S. 2003, \mnras, 344, 1000

\bibitem[{{Chabrier}(2003)}]{chabrier2003}
{Chabrier}, G. 2003, \pasp, 115, 763

\bibitem[{{Chastenet} {et~al.}(2022){Chastenet}, {Leroy}, \&
  {Schinnerer}}]{CHASTENET22JWST}
{Chastenet}, J., {Leroy}, A., \& {Schinnerer}, E. 2022, \apjl

\bibitem[{{Chastenet} {et~al.}(2019){Chastenet}, {Sandstrom}, {Chiang},
  {Leroy}, {Utomo}, {Bot}, {Gordon}, {Draine}, {Fukui}, {Onishi}, \&
  {Tsuge}}]{chastenet2019}
{Chastenet}, J., {Sandstrom}, K., {Chiang}, I.-D., {et~al.} 2019, \apj, 876, 62

\bibitem[{{Dale} {et~al.}(2009){Dale}, {Cohen}, {Johnson}, {Schuster},
  {Calzetti}, {Engelbracht}, {Gil de Paz}, {Kennicutt}, {Lee}, {Begum},
  {Block}, {Dalcanton}, {Funes}, {Gordon}, {Johnson}, {Marble}, {Sakai},
  {Skillman}, {van Zee}, {Walter}, {Weisz}, {Williams}, {Wu}, \&
  {Wu}}]{dale2009}
{Dale}, D.~A., {Cohen}, S.~A., {Johnson}, L.~C., {et~al.} 2009, \apj, 703, 517

\bibitem[{{DeFrees} \& {Miller}(1989)}]{defrees1989}
{DeFrees}, D.~J., \& {Miller}, M.~D. 1989, in Interstellar Dust, ed. L.~J.
  {Allamandola} \& A.~G.~G.~M. {Tielens}, Vol. 135, 173--176

\bibitem[{{Deger} {et~al.}(2022){Deger}, {Lee}, {Whitmore}, {Thilker},
  {Boquien}, {Chandar}, {Dale}, {Ubeda}, {White}, {Grasha}, {Glover},
  {Schruba}, {Barnes}, {Klessen}, {Kruijssen}, {Rosolowsky}, \&
  {Williams}}]{deger2022}
{Deger}, S., {Lee}, J.~C., {Whitmore}, B.~C., {et~al.} 2022, \mnras, 510, 32

\bibitem[{{Draine} \& {Li}(2001)}]{draine2001}
{Draine}, B.~T., \& {Li}, A. 2001, \apj, 551, 807

\bibitem[{{Draine} {et~al.}(2021){Draine}, {Li}, {Hensley}, {Hunt},
  {Sandstrom}, \& {Smith}}]{draine2021}
{Draine}, B.~T., {Li}, A., {Hensley}, B.~S., {et~al.} 2021, \apj, 917, 3

\bibitem[{{Draine} {et~al.}(2007){Draine}, {Dale}, {Bendo}, {Gordon}, {Smith},
  {Armus}, {Engelbracht}, {Helou}, {Kennicutt}, {Li}, {Roussel}, {Walter},
  {Calzetti}, {Moustakas}, {Murphy}, {Rieke}, {Bot}, {Hollenbach}, {Sheth}, \&
  {Teplitz}}]{draine2007}
{Draine}, B.~T., {Dale}, D.~A., {Bendo}, G., {et~al.} 2007, \apj, 663, 866

\bibitem[{{Egorov} {et~al.}(2022){Egorov}, {Leroy}, \&
  {Schinnerer}}]{EGOROV22JWST}
{Egorov}, D., {Leroy}, A., \& {Schinnerer}, E. 2022, \apjl

\bibitem[{{Emsellem} {et~al.}(2022){Emsellem}, {Schinnerer}, {Santoro},
  {Belfiore}, {Pessa}, {McElroy}, {Blanc}, {Congiu}, {Groves}, {Ho}, {Kreckel},
  {Razza}, {Sanchez-Blazquez}, {Egorov}, {Faesi}, {Klessen}, {Leroy}, {Meidt},
  {Querejeta}, {Rosolowsky}, {Scheuermann}, {Anand}, {Barnes},
  {Be{\v{s}}li{\'c}}, {Bigiel}, {Boquien}, {Cao}, {Chevance}, {Dale},
  {Eibensteiner}, {Glover}, {Grasha}, {Henshaw}, {Hughes}, {Koch}, {Kruijssen},
  {Lee}, {Liu}, {Pan}, {Pety}, {Saito}, {Sandstrom}, {Schruba}, {Sun},
  {Thilker}, {Usero}, {Watkins}, \& {Williams}}]{emsellem2022}
{Emsellem}, E., {Schinnerer}, E., {Santoro}, F., {et~al.} 2022, \aap, 659, A191

\bibitem[{{Engelbracht} {et~al.}(2005){Engelbracht}, {Gordon}, {Rieke},
  {Werner}, {Dale}, \& {Latter}}]{engelbracht2005}
{Engelbracht}, C.~W., {Gordon}, K.~D., {Rieke}, G.~H., {et~al.} 2005, \apjl,
  628, L29

\bibitem[{{Engelbracht} {et~al.}(2008){Engelbracht}, {Rieke}, {Gordon},
  {Smith}, {Werner}, {Moustakas}, {Willmer}, \& {Vanzi}}]{engelbracht2008}
{Engelbracht}, C.~W., {Rieke}, G.~H., {Gordon}, K.~D., {et~al.} 2008, \apj,
  678, 804

\bibitem[{{Galliano} {et~al.}(2021){Galliano}, {Nersesian}, {Bianchi}, {De
  Looze}, {Roychowdhury}, {Baes}, {Casasola}, {Cassar{\'a}}, {Dobbels},
  {Fritz}, {Galametz}, {Jones}, {Madden}, {Mosenkov}, {Xilouris}, \&
  {Ysard}}]{galliano2021}
{Galliano}, F., {Nersesian}, A., {Bianchi}, S., {et~al.} 2021, \aap, 649, A18

\bibitem[{{Hassani} {et~al.}(2022){Hassani}, {Leroy}, \&
  {Schinnerer}}]{HASSANI22JWST}
{Hassani}, H., {Leroy}, A., \& {Schinnerer}, E. 2022, \apjl

\bibitem[{{Hollenbach} \& {Tielens}(1997)}]{hollenbach1997}
{Hollenbach}, D.~J., \& {Tielens}, A.~G.~G.~M. 1997, \araa, 35, 179

\bibitem[{{Jensen} {et~al.}(2017){Jensen}, {H{\"o}nig}, {Rakshit},
  {Alonso-Herrero}, {Asmus}, {Gandhi}, {Kishimoto}, {Smette}, \&
  {Tristram}}]{jensen2017}
{Jensen}, J.~J., {H{\"o}nig}, S.~F., {Rakshit}, S., {et~al.} 2017, \mnras, 470,
  3071

\bibitem[{{Kourkchi} \& {Tully}(2017)}]{kourkchi2017}
{Kourkchi}, E., \& {Tully}, R.~B. 2017, \apj, 843, 16

\bibitem[{{Lai} {et~al.}(2020){Lai}, {Smith}, {Baba}, {Spoon}, \&
  {Imanishi}}]{lai2020}
{Lai}, T. S.~Y., {Smith}, J.~D.~T., {Baba}, S., {Spoon}, H. W.~W., \&
  {Imanishi}, M. 2020, \apj, 905, 55

\bibitem[{{Lee} {et~al.}(2022{\natexlab{a}}){Lee}, {Leroy}, \&
  {Schinnerer}}]{LEE22JWST}
{Lee}, J., {Leroy}, A., \& {Schinnerer}, E. 2022{\natexlab{a}}, \apjl

\bibitem[{{Lee} {et~al.}(2022{\natexlab{b}}){Lee}, {Whitmore}, {Thilker},
  {Deger}, {Larson}, {Ubeda}, {Anand}, {Boquien}, {Chandar}, {Dale},
  {Emsellem}, {Leroy}, {Rosolowsky}, {Schinnerer}, {Schmidt}, {Lilly},
  {Turner}, {Van Dyk}, {White}, {Barnes}, {Belfiore}, {Bigiel}, {Blanc}, {Cao},
  {Chevance}, {Congiu}, {Egorov}, {Glover}, {Grasha}, {Groves}, {Henshaw},
  {Hughes}, {Klessen}, {Koch}, {Kreckel}, {Kruijssen}, {Liu}, {Lopez},
  {Mayker}, {Meidt}, {Murphy}, {Pan}, {Pety}, {Querejeta}, {Razza}, {Saito},
  {S{\'a}nchez-Bl{\'a}zquez}, {Santoro}, {Sardone}, {Scheuermann}, {Schruba},
  {Sun}, {Usero}, {Watkins}, \& {Williams}}]{lee2022}
{Lee}, J.~C., {Whitmore}, B.~C., {Thilker}, D.~A., {et~al.} 2022{\natexlab{b}},
  \apjs, 258, 10

\bibitem[{{Leroy} {et~al.}(2022){Leroy}, {Schinnerer}, \&
  {Rosolowsky}}]{LEROY22JWST}
{Leroy}, A., {Schinnerer}, E., \& {Rosolowsky}, E. 2022, \apjl

\bibitem[{{Leroy} {et~al.}(2021){Leroy}, {Schinnerer}, {Hughes}, {Rosolowsky},
  {Pety}, {Schruba}, {Usero}, {Blanc}, {Chevance}, {Emsellem}, {Faesi},
  {Herrera}, {Liu}, {Meidt}, {Querejeta}, {Saito}, {Sandstrom}, {Sun},
  {Williams}, {Anand}, {Barnes}, {Behrens}, {Belfiore}, {Benincasa},
  {Be{\v{s}}li{\'c}}, {Bigiel}, {Bolatto}, {den Brok}, {Cao}, {Chandar},
  {Chastenet}, {Chiang}, {Congiu}, {Dale}, {Deger}, {Eibensteiner}, {Egorov},
  {Garc{\'\i}a-Rodr{\'\i}guez}, {Glover}, {Grasha}, {Henshaw}, {Ho}, {Kepley},
  {Kim}, {Klessen}, {Kreckel}, {Koch}, {Kruijssen}, {Larson}, {Lee}, {Lopez},
  {Machado}, {Mayker}, {McElroy}, {Murphy}, {Ostriker}, {Pan}, {Pessa},
  {Puschnig}, {Razza}, {S{\'a}nchez-Bl{\'a}zquez}, {Santoro}, {Sardone},
  {Scheuermann}, {Sliwa}, {Sormani}, {Stuber}, {Thilker}, {Turner}, {Utomo},
  {Watkins}, \& {Whitmore}}]{leroy2021}
{Leroy}, A.~K., {Schinnerer}, E., {Hughes}, A., {et~al.} 2021, \apjs, 257, 43

\bibitem[{{Madden} {et~al.}(2006){Madden}, {Galliano}, {Jones}, \&
  {Sauvage}}]{madden2006}
{Madden}, S.~C., {Galliano}, F., {Jones}, A.~P., \& {Sauvage}, M. 2006, \aap,
  446, 877

\bibitem[{{Maragkoudakis} {et~al.}(2020){Maragkoudakis}, {Peeters}, \&
  {Ricca}}]{maragkoudakis2020}
{Maragkoudakis}, A., {Peeters}, E., \& {Ricca}, A. 2020, \mnras, 494, 642

\bibitem[{{Press}(1993)}]{press1993}
{Press}, W.~H. 1993, Science, 259, 1931

\bibitem[{{R{\'e}my-Ruyer} {et~al.}(2015){R{\'e}my-Ruyer}, {Madden},
  {Galliano}, {Lebouteiller}, {Baes}, {Bendo}, {Boselli}, {Ciesla}, {Cormier},
  {Cooray}, {Cortese}, {De Looze}, {Doublier-Pritchard}, {Galametz}, {Jones},
  {Karczewski}, {Lu}, \& {Spinoglio}}]{remyruyer2015}
{R{\'e}my-Ruyer}, A., {Madden}, S.~C., {Galliano}, F., {et~al.} 2015, \aap,
  582, A121

\bibitem[{{Riechers} {et~al.}(2014){Riechers}, {Pope}, {Daddi}, {Armus},
  {Carilli}, {Walter}, {Hodge}, {Chary}, {Morrison}, {Dickinson},
  {Dannerbauer}, \& {Elbaz}}]{riechers2014}
{Riechers}, D.~A., {Pope}, A., {Daddi}, E., {et~al.} 2014, \apj, 786, 31

\bibitem[{{Rieke} {et~al.}(2015){Rieke}, {Wright}, {B{\"o}ker}, {Bouwman},
  {Colina}, {Glasse}, {Gordon}, {Greene}, {G{\"u}del}, {Henning}, {Justtanont},
  {Lagage}, {Meixner}, {N{\o}rgaard-Nielsen}, {Ray}, {Ressler}, {van Dishoeck},
  \& {Waelkens}}]{rieke2015}
{Rieke}, G.~H., {Wright}, G.~S., {B{\"o}ker}, T., {et~al.} 2015, \pasp, 127,
  584

\bibitem[{{Rieke} {et~al.}(2005){Rieke}, {Kelly}, \& {Horner}}]{rieke2005}
{Rieke}, M.~J., {Kelly}, D., \& {Horner}, S. 2005, in Society of Photo-Optical
  Instrumentation Engineers (SPIE) Conference Series, Vol. 5904, Cryogenic
  Optical Systems and Instruments XI, ed. J.~B. {Heaney} \& L.~G. {Burriesci},
  1--8

\bibitem[{{Rigopoulou} {et~al.}(2021){Rigopoulou}, {Barale}, {Clary}, {Shan},
  {Alonso-Herrero}, {Garc{\'\i}a-Bernete}, {Hunt}, {Kerkeni},
  {Pereira-Santaella}, \& {Roche}}]{rigopoulou2021}
{Rigopoulou}, D., {Barale}, M., {Clary}, D.~C., {et~al.} 2021, \mnras, 504,
  5287

\bibitem[{{Sandstrom} {et~al.}(2022){Sandstrom}, {Leroy}, \&
  {Schinnerer}}]{SANDSTROM22JWST}
{Sandstrom}, K., {Leroy}, A., \& {Schinnerer}, E. 2022, \apjl

\bibitem[{{Sandstrom} {et~al.}(2010){Sandstrom}, {Bolatto}, {Draine}, {Bot}, \&
  {Stanimirovi{\'c}}}]{sandstrom2010}
{Sandstrom}, K.~M., {Bolatto}, A.~D., {Draine}, B.~T., {Bot}, C., \&
  {Stanimirovi{\'c}}, S. 2010, \apj, 715, 701

\bibitem[{{Shaya} {et~al.}(2017){Shaya}, {Tully}, {Hoffman}, \&
  {Pomar{\`e}de}}]{shaya2017}
{Shaya}, E.~J., {Tully}, R.~B., {Hoffman}, Y., \& {Pomar{\`e}de}, D. 2017,
  \apj, 850, 207

\bibitem[{{Smith} {et~al.}(2007){Smith}, {Draine}, {Dale}, {Moustakas},
  {Kennicutt}, {Helou}, {Armus}, {Roussel}, {Sheth}, {Bendo}, {Buckalew},
  {Calzetti}, {Engelbracht}, {Gordon}, {Hollenbach}, {Li}, {Malhotra},
  {Murphy}, \& {Walter}}]{smith2007}
{Smith}, J.~D.~T., {Draine}, B.~T., {Dale}, D.~A., {et~al.} 2007, \apj, 656,
  770

\bibitem[{{Thilker} {et~al.}(2022){Thilker}, {Whitmore}, {Lee}, {Deger},
  {Chandar}, {Larson}, {Hannon}, {Ubeda}, {Dale}, {Glover}, {Grasha},
  {Klessen}, {Kruijssen}, {Rosolowsky}, {Schruba}, {White}, \&
  {Williams}}]{thilker2022}
{Thilker}, D.~A., {Whitmore}, B.~C., {Lee}, J.~C., {et~al.} 2022, \mnras, 509,
  4094

\bibitem[{{Tielens}(2008)}]{tielens2008}
{Tielens}, A.~G.~G.~M. 2008, \araa, 46, 289

\bibitem[{{Turner} {et~al.}(2021){Turner}, {Dale}, {Lee}, {Boquien}, {Chandar},
  {Deger}, {Larson}, {Mok}, {Thilker}, {Ubeda}, {Whitmore}, {Belfiore},
  {Bigiel}, {Blanc}, {Emsellem}, {Grasha}, {Groves}, {Klessen}, {Kreckel},
  {Kruijssen}, {Leroy}, {Rosolowsky}, {Sanchez-Blazquez}, {Schinnerer},
  {Schruba}, {Van Dyk}, \& {Williams}}]{turner2021}
{Turner}, J.~A., {Dale}, D.~A., {Lee}, J.~C., {et~al.} 2021, \mnras, 502, 1366

\bibitem[{{Whitmore} {et~al.}(2021){Whitmore}, {Lee}, {Chandar}, {Thilker},
  {Hannon}, {Wei}, {Huerta}, {Bigiel}, {Boquien}, {Chevance}, {Dale}, {Deger},
  {Grasha}, {Klessen}, {Kruijssen}, {Larson}, {Mok}, {Rosolowsky},
  {Schinnerer}, {Schruba}, {Ubeda}, {Van Dyk}, {Watkins}, \&
  {Williams}}]{whitmore2021}
{Whitmore}, B.~C., {Lee}, J.~C., {Chandar}, R., {et~al.} 2021, \mnras, 506,
  5294

\bibitem[{{Wolfire} {et~al.}(2022){Wolfire}, {Vallini}, \&
  {Chevance}}]{wolfire2022}
{Wolfire}, M.~G., {Vallini}, L., \& {Chevance}, M. 2022, \araa, 60, 247

\bibitem[{{Wu} {et~al.}(2006){Wu}, {Charmandaris}, {Hao}, {Brandl},
  {Bernard-Salas}, {Spoon}, \& {Houck}}]{wu2006}
{Wu}, Y., {Charmandaris}, V., {Hao}, L., {et~al.} 2006, \apj, 639, 157

\bibitem[{{Wu} {et~al.}(2010){Wu}, {Helou}, {Armus}, {Cormier}, {Shi}, {Dale},
  {Dasyra}, {Smith}, {Papovich}, {Draine}, {Rahman}, {Stierwalt}, {Fadda},
  {Lagache}, \& {Wright}}]{wu2010}
{Wu}, Y., {Helou}, G., {Armus}, L., {et~al.} 2010, \apj, 723, 895

\end{thebibliography}




\end{document}